\documentclass[12pt]{article}
\pdfoutput=1
\usepackage{mathrsfs}
\usepackage{amsmath}
\usepackage{mathrsfs}
\usepackage{amssymb}
\usepackage{color}
\usepackage[unicode]{hyperref}
\usepackage{graphicx}
\usepackage{xcolor}
\usepackage{hyperref}
\usepackage{color}
\usepackage[utf8]{inputenc}
\usepackage{setspace}
\usepackage{amsmath, amssymb, amsthm}
\numberwithin{equation}{section}
\hypersetup{colorlinks=false, linkcolor=blue, citecolor=red}
\newcommand{\be}{\begin{equation}}
\newcommand{\ee}{\end{equation}}
\def\g{\gamma}
\def\G{\Gamma}

\def\m{\mu}

\def\e{\epsilon}
\def\l{\lambda}

\def\d{\delta}

\def\f{\phi}

\def\D{\Delta}

\def\ta{\tau}

\newcommand{\bea}{\begin{eqnarray}}
\newcommand{\eea}{\end{eqnarray}}

\renewcommand{\d}{\delta}
\newcommand{\dsl}{\pa \kern-0.5em /}
\newcommand{\la}{\lambda}
\newcommand{\half}{\frac{1}{2}}
\newcommand{\pa}{\partial}

\newcommand{\nn}{\nonumber\\}

\begin{document}
	\title{\bf  Towards a Bootstrap approach to higher orders \\ of epsilon expansion}
	\date{}
\author{Parijat Dey${}^{a}$\footnote{parijatdey@iisc.ac.in},~Apratim Kaviraj${}^{a,b}$\footnote{apratim.kaviraj@lpt.ens.fr} \\  \\
	\it ${}^a$Centre for High Energy Physics,
	\it Indian Institute of Science,\\ \it C.V. Raman Avenue, Bangalore 560012, India. \\ 
	\it ${}^b$Laboratoire de Physique Th\'{e}orique \\ \& \it Institut de Physique Th\'{e}orique Philippe Meyer, \\
\it 	\'{E}cole Normale Supérieure, PSL Research University,\\ 
\it 	24 rue Lhomond, 
	75231 Paris Cedex 05, France
    }	
\maketitle
 \abstract{We employ a hybrid approach in determining the anomalous dimension and OPE coefficient of higher spin  operators in the  Wilson-Fisher theory. First we do a large spin analysis for CFT data where we use results obtained from the usual and the Mellin Bootstrap and also from Feynman diagram literature. This gives new predictions at $O(\e^4)$  and $O(\e^5)$ for anomalous dimensions and OPE coefficients, and also provides a cross-check for the results from  Mellin Bootstrap. These higher orders get contributions from all higher spin operators in the crossed channel. We also use the Bootstrap in Mellin  space method for $\phi^3$ in $d=6-\e$ CFT  where we calculate general higher spin OPE data. We  demonstrate a higher loop order calculation in this approach by summing over contributions from higher spin operators of the crossed channel in the same spirit as before.}
\tableofcontents

\onehalfspacing

\section{Introduction}
Conformal Field theories (CFT) are interesting to Physicists for their roles in critical points of phase transition and studying RG flows. In the study of CFTs it is important to understand how to determine the operator spectrum of a theory. The conformal bootstrap is an approach, introduced in the pioneering works of \cite{migdal, fer, polya, bpz}   that is solely based on using the symmetries of a CFT. The last decade has seen a revival of the conformal bootstrap program following the work of \cite{rrtv} . This involved the equality (crossing symmetry) of the Operator Product Expansion (OPE) in the direct channel and crossed channel. It has been very successful in obtaining numerical results, in theories like Ising model \cite{3dising, mostprecise} . 
 The idea here is to constrain the space of CFTs starting with some given assumptions which are usually based on  symmetries or unitarity of the theory (see \cite{others}-\cite{sleight} for related works.)

Another goal of the boostrap program is to solve a CFT analytically. One aspect is using the bootstrap equation in a  lightcone limit, introduced in \cite{anboot, komargodski} . This assumes a higher spin sector of operators in a theory, whose anomalous dimensions and OPE coefficients can be computed in terms of a lower twist operators present in the spectrum. There has been many subsequent works \cite{alday1}-\cite{kss2} , that has taken this approach further. In these works a systematic approach has been developed to compute the anomalous dimension of large spin double trace operators as an asymptotic expansion in inverse spin. . 

This brings us to the other aspect of analytic approach, which is to give an alternative way to Feynman diagrams, to calculate the OPE spectrum in CFTs with a perturbative parameter. In perturbative CFTs, such as the Wilson-Fisher fixed point in $\phi^4$ in $d=4-\e$, Gross-Neveu model \cite{rychkovtan}   have shown various techniques of obtaining OPE data under expansion in a small parameter. The twist conformal blocks can be used efficiently to extract the leading order anomalous dimension in $4-\e$ dimensions\cite{alday3}. In \cite{sen}  a dispersion relation-based technique was used, that was inspired from the original work of Polyakov\cite{polya} .

A new approach to bootstrap, also based on Polykov's work \cite{polya}, was chalked out in \cite{prl, uslong} . This involved replacing conformal blocks with a manifestly crossing symmetric basis of Witten diagrams in the Operator Product Expansion. The use of Witten diagrams requires the description to be in Mellin space. Consistency with the usual OPE then requires satisfaction of certain equations. These equations, referred to as  Bootstrap equations in Mellin space, can be used to solve OPE data order by order in a perturbative parameter. It gave many results for the Wilson-Fisher model and was generalized for global symmetries in \cite{booton} . The method was also successful in reproducing the large spin results in lightcone bootstrap, treating the inverse of spin as a small parameter. The equivalence of this new approach to bootstrap with usual one has been studied recently in \cite{usnew} . 

All the analytic approaches mentioned above reach their limits at  certain orders in perturbation. The large spin literature gives a systematic expansion in all orders of large spin. These results have then been used to obtain higher spin anomalous dimensions perturbatively, for example in Wilson-Fisher up to $O(\e^2)$ order \cite{alday3} . The Mellin bootstrap equations are successful in giving  both higher spin OPE coefficients and anomalous dimensions up to $O(\e^3)$, with significant ease \cite{prl, uslong, booton}. There are certain complications that arise beyond this order which makes it difficult to get higher loops results. Even though such difficulties are expected at higher orders, it is desirable to know how much can be done with the present bootstrap-driven techniques and without further intricacy.

The goal of this paper is to present some calculations for the higher spin double trace operators with the tools of the known methods. The calculations are simple but will take us to some high orders in perturbation. The results presented are mostly unknown in the Feynman diagram literature. In the first half of the paper we have used the large spin analysis from the usual bootstrap approach to compute anomalous dimensions and OPE coefficients of up to $O(\e^5)$ and $O(\e^4)$ respectively for the Wilson-Fisher theory in $d=4-\e$. This computation takes in information of infinite higher spin minimal twist operators.  This comes from existing $\e$-expansion results obtained with Mellin Bootstrap as well as those from Feynman diagram literature. In \cite{usnew} it is shown that the difference between the usual and Mellin bootstrap starts at $O(\g_{\ell}^2)$ for the double trace operators having dimensions $\D=d-2+\ell+\g_{\ell}$. For higher spin operators in $\f^4$ theory, $\g_{\ell} \sim \frac{1}{\ell^2}$ in the large spin limit. Hence the large spin expressions from usual and Mellin approach should agree uptil $O(1/\ell^4)$. Since we calculate up to $O(\e^5/\ell^3)$ we can safely use these formulae.  The second half of the paper uses the ideas of Mellin bootstrap. The theory used is $\phi^3$ theory in $d=6-\e$, for convenience. Here we compute OPE data at one loop. We also present a calculation that takes us to a higher order in $\e$. The main objective of this part is to understand how Mellin bootstrap ideas can be used to systematically compute higher loop results.

The paper is organized as follows. In section  \ref{epexp} we begin by the computation of the anomalous dimension and OPE coeffient in the large spin limit  for the $\f^4$ theory in $4-\e$ dimension using the known results from large spin analytic bootstrap. Section \ref{ep6d} is dedicated to the study of $\f^3$ theory in $6-\e$ dimensions using the ideas of Mellin Bootstrap. We conclude in section \ref{disc} with a brief discussion of the future directions. The appendices  give the calculational details of the paper.

\section{Higher orders of Wilson-Fisher from large spin }\label{epexp}
In this section we derive the anomalous dimensions and OPE coefficients of the operators in $\e$ expansion for  $\f^4$ theory in $4-\e$ dimension in the large spin limit. 

Let us take the OPE of the scalars $\phi \times \phi$. We know that the operator content of this OPE consists of higher spin double-field operators of the schematic form,
\be\label{hsop}
O_{2m,\ell}\sim \phi (\partial^2)^m \partial_{\mu_1}\cdots\partial_{\mu_\ell}\phi\,.
\ee
with conformal dimension,
\be\label{dimdef}
\D=2\D_\phi+2m+\ell+\g \,.
\ee
We consider such operators at large spin, for which it was shown in \cite{anboot, komargodski} that the leading anomalous dimension is determined from the operator(s) having the minimum nonzero twist $\tau_m$. If we assume that the anomalous dimension at large spin has the following expansion,
\be
\g= \g_0 +\frac{\g_1}{\ell}+ \cdots\,,
\ee
then we have\cite{anboot, komargodski, alday1, usnew}, 
\begin{align}\label{anmdim}
\g_0 & = \sum_{\ell_m} -C_{\ell_m} \frac{2 \Gamma^2 \left(\Delta _{\phi }\right)\,  \Gamma \left(2 \ell _m+\tau _m\right)}{\Gamma \left(\ell _m+\frac{\tau _m}{2}\right){}^2 \Gamma \left(\Delta _{\phi }-\frac{\tau _m}{2}\right){}^2}\left(\frac{1}{\ell }\right)^{\tau _m}\,,\nn
\g_1& = \sum_{\ell_m} C_{\ell_m}\, \frac{\left(2 \Delta _{\phi }-1\right) \tau _m \Gamma^2 \left(\Delta _{\phi }\right)\, 2^{\tau _m+2 \ell _m-1}  \Gamma \left(\ell _m+\frac{\tau _m}{2}+\frac{1}{2}\right)}{\sqrt{\pi } \Gamma \left(\ell _m+\frac{\tau _m}{2}\right) \Gamma^2 \left(\Delta _{\phi }-\frac{\tau _m}{2}\right){}} \left(\frac{1}{\ell }\right)^{\tau _m}\,.
\end{align}
Note that there is a sum over $\ell_m$ in case there are multiple operators with the same minimal twist $\tau_m$. The OPE coefficient of each operator is given by $C_{\ell_m}$. 

In a similar way, the OPE coefficients of these operators at large spin, can be expressed in terms of the minimal twist operator(s) as follows,
\be
\delta C_{\ell}= C^{(0)}_{\ell}\bigg(1+\delta C_0+\frac{\delta C_1}{\ell}+\cdots\bigg)\,,
\ee
where we have,
\begin{align}\label{ope}
\delta C_0 &=-\sum_{\ell_m}C_{\ell_m}\,\frac{2 \Gamma^2 \left(\Delta _{\phi }\right) \Gamma \left(\tau _m+2 \ell_m\right) }{\Gamma^2 \left(\frac{\tau _m}{2}+\ell_m\right) \Gamma^2 \left(\Delta _{\phi }-\frac{\tau _m}{2}\right)}\,\bigg(\psi \left(\ell_m+\frac{\tau _m}{2}\right)+\gamma_E -\log (2)\bigg)\left(\frac{1}{\ell }\right)^{\tau _m}\,,\nn
\delta C_1 &=\sum_{\ell_m}C_{\ell_m}\,\frac{\Gamma^2 \left(\Delta _{\phi }\right){}\, \Gamma \left(\tau _m\right) }{2 \Gamma^2 \left(\frac{\tau _m}{2}\right)\, \Gamma^2 \left(\Delta _{\phi }-\frac{\tau _m}{2}\right)}\,\bigg(-2 \Delta _{\phi } \left(\log (4) \tau _m+2\right)+2 \left(2 \Delta _{\phi }-1\right) \tau _m \left(\psi \left(\frac{\tau _m}{2}\right)+\gamma_E \right) \nn & \hspace{6cm} +(2+\log (4)) \tau _m+3\bigg)\left(\frac{1}{\ell }\right)^{\tau _m}\,.
\end{align}
The subsequent orders of $\g$ and $\d C_{\ell}$ in $1/\ell$ can also be computed easily using the techniques in \cite{alday1, usnew}. However for simplicity we will focus only on the leading order terms.
\subsection{Anomalous dimension}
This section deals with the $\e$ expansion of $\phi^4$ theory in $d=4-\e$ dimensions (Wilson-Fisher fixed point). We will use the above formulas to get the OPE data of double field operators at large spin. The dimension of the fundamental scalar $\f$ reads 
\be
\D_\f = 1-\half\, \e+\frac{1}{108}\,\e^2+\frac{109}{11664}\,\e^3+\d_\f^{(4)}\,\e^4+\d_\f^{(5)}\,\e^5 + O(\e^6)\,.
\ee
For this theory the minimal twist operators are the double-field operators $\phi\partial^\ell \phi$ themselves. This is because under the $\e$-expansion (eq \eqref{dimdef}), their twists are $\tau_m=\tau=2+O(\e)$\,.
For $\ell_m=0$ we have the scalar $\phi^2$ operator. Its twist and OPE coefficient are  respectively given by,
\begin{align}\label{scalar2}
 \tau_m &= 2-\frac{2}{3}\,\e+\frac{19}{162}\,\e^2+ \d_0^{(3)}\,\e^3+ \d_0^{(4)}\,\e^4+ \d_0^{(5)}\,\e^5 + O(\e^6)\,,\nn
 C_0 &= 2-\frac{2 \epsilon }{3}-\frac{34 \epsilon ^2}{81}+ C_0^{(3)} \epsilon ^3+ C_0^{(4)} \epsilon ^4+ C_0^{(5)} \epsilon ^5+ O(\e^6)\,.
\end{align}
Substituting \eqref{scalar2} in the first term of \eqref{anmdim} for $\ell_m=0$ we get,
\begin{align}\label{c1}
\g_0|_{\ell_m=0} & = -\frac{\e^2}{9\,\ell^2}+\frac{\epsilon ^3 (-18 \log (\ell )-18 \gamma_E +11)}{243 \ell ^2}\nn
&+\frac{\epsilon ^4 \left(-17496\, {\d_0^{(3)}}-162 \log (\ell ) (4 \log (\ell )+8 \gamma_E -7)-162 \pi ^2+162 \gamma_E  (7-4 \gamma_E )+1421\right)}{26244 \ell ^2}\nn & -\frac{\epsilon ^5}{472392 \ell ^2}\bigg(-629856 \,{\d_\f^{(4)}}+72 \log (\ell ) \left(2187 {\d_0^{(3)}}+3 \log (\ell ) (12 \log (\ell )+36 \gamma_E -41)\right.\nn & \left.+27 \pi ^2+6 \gamma_E  (18 \gamma_E -41)-202\right)+5832 (27 \gamma_E -29) {\d_0^{(3)}}+314928 {\d_0^{(4)}}+26244\, {C_0^{(3)}}+8100 \zeta (3)\nn &-1755 \pi ^2+72 \gamma_E  \left(3 \gamma_E  (12 \gamma_E -41)+27 \pi ^2-202\right)+8858\bigg) +O(\e^6)\,.
\end{align}


Here $\g_0|_{\ell_m=0}$ is what one gets from \eqref{anmdim} with only $\ell_m=0$.

Now let us consider the contribution from higher spin minimal twist operators of \eqref{hsop}. Note that the operators $O_{2m,\ell}$ with $m>0$ can contribute at orders suppressed as $\ell^{-4}$ or beyond, and hence would not contribute at $\ell^{-2}$ or $\ell^{-3}$, which we consider in this paper.  
It is the same for other higher twist operators (like those composed of four or more $\phi$-s) too. For $O_{0,\ell}=\phi\partial^\ell\phi$, the twists and OPE coefficients are respectively,
\begin{align}\label{hs}
C_{\ell_m} &= \frac{2 \left(\ell _m!\right){}^2}{\left(2 \ell _m\right)!}-\frac{2 \epsilon  \left(\ell _m!\right){}^2 \left(2 H_{\ell _m}-H_{2 \ell _m}\right)}{\left(2 \ell _m\right)!}+C_{\ell_m}^{(2)}\,\e^2+ C_{\ell_m}^{(3)}\,\e^3+ O(\e^4),\nn
 \tau_m &= 2-\e+ \frac{1}{54} \epsilon ^2 \left(1-\frac{6}{\ell _m \left(\ell _m+1\right)}\right)\nn
& +\frac{\epsilon ^3 \left(-432 \left(\ell _m+1\right) \ell _m H_{\ell _m}+109 \left(\ell _m+2\right) \ell _m^3+373 \ell _m^2-384 \ell _m-324\right)}{5832 \ell _m^2 \left(\ell _m+1\right){}^2} + O(\e^4)\,.
\end{align}
Substituting \eqref{hs} in \eqref{anmdim} we obtain  the following,
\begin{align}\label{hspin}
\g_0|_{\ell_m>0} & =\sum_{\ell_m>0}\bigg(-\e^4\,\frac{1+2\ell_m}{81(\ell_m+\ell^2_m)^2\,\ell^2} -\frac{\epsilon ^5}{2187 \ell ^2 \ell _m^3 \left(\ell _m+1\right){}^3}\nn & \times \left(27+9 \ell _m \left(\ell _m+1\right) \left(2 \ell _m+1\right) \left(H_{\ell _m}+3 \log (\ell )\right)+\right.\left.\ell _m \left(\ell _m+1\right) \left((54 \gamma_E -44) \ell _m+27 \gamma_E +59\right)\right)\bigg)\,.
\end{align}
Here $\g_0|_{\ell_m>0}$ is the contribution of all spins $\ell_m>0$ to $\g_0$. This sum is over even spins $\ell_m$ and can be performed exactly. For the $O(\e^4)$ term it gives,
\be
\sum_{\ell_m=2}^{\infty} - \frac{1+2\ell_m}{81(\ell_m+\ell^2_m)^2} =\frac{\pi^2-12}{972}\,.
\ee
Hence the the anomalous dimension at $O(\e^4)$  in the large $\ell$ limit is given by,
\be
\g_{0}|_{O(\e^4)} \sim \e^4 \bigg(\frac{-17496\,{\d_0^{(3)}}-162 \log (\ell ) (4 \log (\ell )+8 \gamma_E -7)-162 \pi ^2+162 \gamma_E  (7-4 \gamma_E )+1421}{26244 \ell ^2}+\frac{\pi^2-12}{972\,\ell^2}\bigg)
\ee
Now if we take the $\e^3$ dimension of $\f^2$ as an external input i.e. $\d_0^{(3)}=\frac{937}{17496}-\frac{4 \zeta (3)}{27}$ \cite{klein} (computed using Feynman diagrams), then the anomalous dimension at $O(\e^4)$  in the large $\ell$ limit reads,
\be
\g_{0}|_{O(\e^4)} \sim \frac{\epsilon ^4}{26244 \ell ^2}\bigg(2592 \zeta (3)-162 \log (\ell ) (4 \log (\ell )+8 \gamma_E -7)+162 \gamma_E  (7-4 \gamma_E )-135 \pi ^2+160\bigg)
\ee
This matches precisely with the $\ell^{-2}$ term known in literature \cite{phi4, manashov2}\footnote{There is a typo in \cite{phi4} and the correct expression can be read off from \cite{manashov2}. 
}.

Now we will compute the $O(\e^5)$ anomalous dmension of these operators in the large $\ell$ limit. In order to do that we need to perform the $\ell_m$ sum in \eqref{hspin} at $O(\e^5)$. We will first focus on the terms in \eqref{hspin} without the Harmonic numbers. This sum (over even spins only) can be easily done resulting in the following expression,
\begin{align}\label{sum1}
& \sum_{\ell_m=2}^{\infty} \frac{1}{2187 \ell ^2 \ell _m^3 \left(\ell _m+1\right){}^3}\bigg(-\ell _m \left(\ell _m+1\right) \left((54 \gamma_E -44) \ell _m+27 \gamma_E +59\right)-27 \ell _m \left(\ell _m+1\right) \left(2 \ell _m+1\right) \log (\ell )-27\bigg)\nn
& =\frac{1}{26244 \ell^2}\left( 27 \left(\pi ^2-12\right) \log (\ell)+243 \,\zeta (3)+27\, \gamma_E  \,\left(\pi ^2-12\right)-22 \pi ^2-60 \right)\,.
\end{align}
The remaining terms in \eqref{hspin} reads,
\begin{align}\label{sum2}
& \sum_{\ell_m=2}^{\infty} -\frac{\left(2 \ell _m+1\right) H_{\ell _m}}{243 \ell ^2 \ell _m^2 \left(\ell _m+1\right){}^2}
=\sum_{\ell_m=2}^{\infty}  -\frac{H_{\ell _m}}{243 \ell ^2 \ell _m^2}+ \frac{H_{\ell _m}}{243 \ell ^2 \left(\ell _m+1\right){}^2}
= -\frac{\zeta (3)}{972 \ell^2}\,.
\end{align}
Adding  \eqref{c1},\eqref{sum1} and \eqref{sum2} we obtain the following contribution at $O(\e^5)$,  
\begin{align}\label{e5anm}
\g_0|_{O(\e^5)} \sim & \frac{\e^5}{1417176 \ell ^2}\left(-27 \gamma_E  \left(24 \gamma_E  (12 \gamma_E -41)+162 \pi ^2-31\right)+4077 \pi ^2+8357\right.\nn
&-27\log (\ell )\left(-31+48 \gamma_E  (18 \gamma_E -41)+162 \pi ^2+\right.24 \log (\ell ) (12 \log (\ell )+36 \gamma_E -41))\nn
& \left.+1889568 \d_{\f}^{(4)} + 162 \zeta (3) (432 (\log (\ell )+\gamma_E )-749)-944784 \,\d_0^{(4)}\right)\,.
\end{align}
Here  we have used the $\e^3$ order OPE coefficient of $\f^2$ i.e. $C_0^{(3)}= \frac{23 \zeta (3)}{54}-\frac{611}{4374}$ which can be computed using the $\e^3$ anomalous dimension of $\f^2$ as the external input \cite{uslong}.
Now  we take the values of $\d_0^{(4)}$ and $\d_\f^{(4)}$ as an external input from \cite{klein, phi4klein}
\be\label{b1c2}
\d_0^{(4)}=-\frac{119 \zeta (3)}{1458}+\frac{40 \zeta (5)}{81}-\frac{\pi ^4}{810}+\frac{24857}{1889568}, \qquad \d_\f^{(4)}= -\frac{2 \zeta (3)}{243}+\frac{7217}{1259712}\,.
\ee
Plugging these values in \ref{e5anm} we obtain the large spin anomalous dimension at $O(\e^5)$,
\begin{align}
\g_0|_{O(\e^5)} \sim & \frac{\epsilon ^5}{7085880 \ell ^2}\bigg(7290 \zeta (3) (48 \log (\ell )+48 \gamma_E -41)-2332800 \zeta (5)\nn &-135 \log (\ell ) \left(24 \log (\ell ) (12 \log (\ell )+36 \gamma_E -41)+48 \gamma_E  (18 \gamma_E -41)+162 \pi ^2-31\right)\nn &+27 \left(5 \gamma_E  (24 \gamma_E  (41-12 \gamma_E )+31)+216 \pi ^4-810 \gamma_E  \pi ^2\right)+20385 \pi ^2+33770\bigg)\,.
\end{align}
Now we will consider the  anomalous dimension term subleading in $\ell$,
\be\label{dimsub}
\g_1= \sum_{\ell_m}C_{\ell_m}\,2^{2 \ell _m+{\tau_m}-1}\frac{(2 \Delta \phi -1) {\tau_m} \Gamma^2 (\Delta \phi ) \,  \Gamma \left(\frac{{\tau_m}}{2}+\ell _m+\frac{1}{2}\right)}{\sqrt{\pi }\, \Gamma^2 \left(\Delta \phi -\frac{{\tau_m}}{2}\right)\, \Gamma \left(\frac{{\tau_m}}{2}+\ell _m\right)}\ell ^{-{\tau_m}-1}
\ee
Substituting \eqref{scalar2} in the \eqref{dimsub} for $\ell_m=0$ we get,
\begin{align}\label{hsgg2}
\frac{\g_1}{\ell}\bigg|_{\ell_m=0}&=\frac{\epsilon ^2}{9 \ell ^3}+\frac{\epsilon ^3 (18 \log (\ell )+18 \gamma_E -47)}{243 \ell ^3}\nn
& +\frac{\epsilon ^4}{26244 \ell ^3}\bigg(-2592 \zeta (3)+162 \log (\ell ) (4 \log (\ell )+8 \gamma_E -23)+162 \gamma_E  (4 \gamma_E -23)+162 \pi ^2+2297\bigg)\nn
& +\frac{\e^5}{2834352 \ell ^3}\bigg(\left(-3779136\, {\d_\f^{(4)}}+1889568\, {\d_0^{(4)}}-33858 \pi ^2+54 \gamma_E  \left(24 \gamma_E  (12 \gamma_E -113)+216 \pi ^2+3941\right)\right.\nn&-12259
 +54 \log (\ell ) \left(24 \log (\ell ) (12 \log (\ell )+36 \gamma_E -113)+48 \gamma_E  (18 \gamma_E -113)+216 \pi ^2+3941\right)\nn
& +324 \zeta (3) (1901-432 (\log (\ell )+\gamma_E )))\bigg)\,.
\end{align}
The higher spin exchanges give,
\begin{align}\label{hsg1}
\frac{\g_1}{\ell}\bigg|_{\ell_m>0}&= \sum_{\ell_m>0}\bigg(\frac{\epsilon ^4 \left(2 \ell _m+1\right)}{81 \ell ^3 \left(\ell _m^2+\ell _m\right){}^2}+ \frac{\epsilon ^5 \left(2 \ell _m+1\right)}{4374 \ell ^3 \ell _m^3 \left(\ell _m+1\right){}^3} \bigg( 54-9\left(H_{\ell _m}-6 H_{2 \ell _m}+3 H_{\ell _m+\frac{1}{2}}\right.\nn
& +\log (64)-6 \log (\ell ))\ell _m(\ell _m+1)+\ell _m \left(-125 \ell _m+54 \gamma_E  \left(\ell _m+1\right)-17\right)\bigg)\bigg)\,.
\end{align}
Adding the $O(\e^4)$ contribution from  \eqref{hsgg2} and \eqref{hsg1} we get,
\be
\frac{\g_1}{\ell}\bigg|_{O(\e^4)} \sim \frac{\e^4}{26244 \ell ^3} \bigg(-2592 \zeta (3)+162 \log (\ell ) (4 \log (\ell )+8 \gamma_E -23)+162 \gamma_E  (4 \gamma_E -23)+135 \pi ^2+2621\bigg)\,. 
\ee

Now we focus on the $O(\e^5)$ term.
The sum in \ref{hsg1} can be done in two steps. First we sum over the terms without the Harmonic numbers. This is given by,
\begin{align}
&\sum_{\ell_m=2}^{\infty}\frac{\epsilon ^5 \left(2 \ell _m+1\right)}{4374 \ell ^3 \ell _m^3 \left(\ell _m+1\right){}^3}\bigg(\ell _m \left(-125 \ell _m+54 \gamma_E  \left(\ell _m+1\right)-17\right)-9 \ell _m \left(\ell _m+1\right) (\log (64)-6 \log (\ell ))+54\bigg)\nn
&= \frac{\epsilon ^5}{52488 \ell ^3}\bigg(-486 \zeta (3)-54 \pi ^2 \log (\ell )+648 \log (2 \ell )-54 \gamma_E  \left(\pi ^2-12\right)-2796+\pi ^2 (233+54 \log (2))\bigg)
\end{align}
Now the terms in \ref{hsg1} with the Harmonic numbers are given by,
\be
\sum_{\ell_m=2}^{\infty}-\frac{\epsilon ^5 \left(2 \ell _m+1\right)}{486 \ell ^3 \ell _m^2 \left(\ell _m+1\right){}^2}\bigg(H_{\ell _m}-6 H_{2 \ell _m}+3 H_{\ell _m+\frac{1}{2}}\bigg)
\ee
We can use the following identity for the Harmonic number,
\begin{align}
H_{2\ell_m} = \half \left(H_{\ell_m}+H_{\ell_m-\half}\right)+\log 2\, {\rm and} \quad
H_{\ell_m+\half}-H_{\ell_m-\half}= \frac{2}{2\ell_m+1}
\end{align}
to write,
\begin{align}
\sum_{\ell_m=2}^{\infty}\frac{\left(2 \ell _m+1\right)}{ \ell _m^2 \left(\ell _m+1\right){}^2}\bigg(H_{\ell _m}-6 H_{2 \ell _m}+3 H_{\ell _m+\frac{1}{2}}\bigg)
&= \sum_{\ell_m=2}^{\infty}\frac{\left(2 \ell _m+1\right)}{ \ell _m^2 \left(\ell _m+1\right){}^2}\bigg(-2H_{\ell_m}-\frac{6(-1+\log 2+2\ell_m \log 2)}{2\ell_m+1}\bigg)\nn
& = -\frac{ \zeta (3)}{2}-18+\frac{1}{2} \pi ^2 (2+\log 2)+6\,\log 2
\end{align}
Hence adding the contribution from \eqref{hsgg2} and \eqref{hsg1} we obtain,
\begin{align}
 \frac{\g_1}{\ell}\bigg|_{O(\e^5)} & \sim 
\frac{\epsilon ^5}{14171760 \ell ^3} \bigg(-10935 \zeta (3) (64 \log (\ell )+64 \gamma_E -211)+4665600 \zeta (5)\nn &+270 \log (\ell ) \left(24 \log (\ell ) (12 \log (\ell )+36 \gamma_E -113)+48 \gamma_E  (18 \gamma_E -113)+162 \pi ^2+4589\right)\nn &+54 \left(5 \gamma_E  (24 \gamma_E  (12 \gamma_E -113)+4589)-216 \pi ^4+810 \gamma_E  \pi ^2\right)-135540 \pi ^2-275305\bigg)
\end{align}
This completes the derivation of the anomalous dimension upto $O(\e^5/\ell^3)$.
\subsection{OPE coefficient}
In this sectin we will compute the large spin correction to the OPE coefficients in $\e$. Substituting \eqref{scalar2} in the first term of \eqref{ope} for $\ell_m=0$ we obtain,
\begin{align}\label{scope}
\delta C_0|_{\ell_m=0} &= \e^2\frac{\log (2)}{9\,\ell^2}+ \frac{\e^3}{486\, \ell^2}\left(36 \log (2) \log (\ell)+3 \pi ^2+(18 \gamma_E -11) \log (4)\right)\nn & + \frac{\epsilon ^4}{52488 \ell ^2} \bigg( \left(4 (81 \gamma_E  (4 \gamma_E -7)-242) \log (2)+27 \pi ^2 (8 \gamma_E -7+\log (4096))\right.\nn &  \left. +648 \zeta (3) (1-8 \log (2))+108 \log (\ell ) \left(12 \log (2) \log (\ell )+2 \pi ^2+(8 \gamma_E -7) \log (8)\right)\right)\bigg)+O(\e^5)\,.
\end{align}
Similarly, the higher spin contribution  \eqref{hs} gives,
\begin{align}\label{hsope}
\delta C_0|_{\ell_m > 0} &= \sum_{\ell_m=2}^{\infty} \frac{\epsilon ^4 \left(2 \ell _m+1\right) \left(\log (2)-H_{\ell _m}\right)}{81 \ell ^2 \ell _m^2 \left(\ell _m+1\right){}^2}
= \e^4\,\frac{21 \zeta (3)-4 \left(\pi ^2-12\right) \log (2)}{3888 \ell ^2}\,.
\end{align}
Adding \eqref{scope} and \eqref{hsope} we finally obtain the correction to the OPE coefficient at $O(\e^4)$,
\begin{align}
\delta C_0\bigg|_{O(\e^4)} \sim &\frac{\e^4}{52488 \ell^2}(4 (81 \gamma_E  (4 \gamma_E -7)-80) \log (2)+27 \pi ^2 (8 \gamma_E -7+10 \log (2))\nn
& +\left.108 \log (\ell) \left(12 \log (2) \log (\ell)+2 \pi ^2+(8 \gamma_E -7) \log (8)\right)+81\zeta (3) (23/2-64 \log (2))\right)\,.
\end{align}
The subleading correction in $\e$ to the OPE coefficients can also be computed easily. We are stopping at $O(\e^4)$ since general spin OPE coefficients are known only till $O(\e^3)$\,.

\subsection{Alternative way of computing scalar OPE coefficient}
One can compute the OPE coefficients of $\phi^2$ using the known results from Feynman diagram as follows. We denote the twist and OPE coefficients of $\f^2$ as,
\begin{align}\label{phi2}
\tau_m &= 2-\frac{2}{3}\,\e+\frac{19}{162}\,\e^2+ \d_0^{(3)}\,\e^3+ \d_0^{(4)}\,\e^4+ \d_0^{(5)}\,\e^5 + O(\e^6)\,,\nn
C_0 &= C_0^{(0)}+ C_0^{(1)}\, \e + C_0^{(2)}\e^2+ C_0^{(3)} \epsilon ^3+ C_0^{(4)} \epsilon ^4+ C_0^{(5)} \epsilon ^5+ O(\e^6)\,.
\end{align}
We would like to compute the $C_0^{i}$'s using the known results of the higher spin anomalous dimension. From the first term of \eqref{anmdim}  we obtain,
\begin{align}\label{gphi}
\g_0 & = - C_0^{(0)} \frac{\epsilon ^2}{18 \ell ^2} -\frac{\epsilon ^3}{486 \ell ^2}\bigg(18 C_0^{(0)} \log (\ell )+2 (9 \gamma_E -1) C_0^{(0)}+27 C_0^{(1)}\bigg)\nn 
&+ \frac{\epsilon ^4}{52488 \ell ^2}\bigg(\left(-17496 \d_0^{(3)}-162 \pi ^2+162 \gamma_E  (3-4 \gamma_E )+881\right) C_0^{(0)}\nn & -162 \log (\ell ) (4 C_0^{(0)} \log (\ell )+(8 \gamma_E -3) C_0^{(0)}+12C_0^{(1)})-108 (2 (9 \gamma_E -1) C_0^{(1)}+27 C_0^{(2)})\bigg)\nn
&+ \frac{\e^4}{\ell^2}\frac{(\pi^2-12)}{972} +O(\e^5)\,.
\end{align}
 Note that the anomalous dimension of the double trace operators $\D$ is known from Feynman diagrams  upto $O(\e^4)$ \cite{phi4, manashov2}, 
\begin{align}\label{gphi2}
\D &= 2-\e+ \frac{1}{54} \epsilon ^2 \left(1-\frac{6}{\ell  \left(\ell +1\right)}\right)+\frac{\epsilon ^3 \left(-432 \left(\ell +1\right) \ell  H_{\ell }+109 \left(\ell +2\right) \ell ^3+373 \ell ^2-384 \ell-324\right)}{5832 \ell^2 \left(\ell +1\right){}^2} \nn
& -\frac{\epsilon ^4}{157464}\bigg(\frac{81}{4} \left(\frac{192 \left(H_{\ell }\right){}^2}{\ell ^2+\ell }-\frac{16 \left(89 \ell ^2+53 \ell -18\right) H_{\ell }}{\ell ^2 (\ell +1)^2}+\frac{288 H_{\ell }^{(2)}}{\ell ^2+\ell }\right.\nn & \left.-\frac{8 \left(\ell ^2+\ell -4\right) \left(3\left(\psi ^{(1)}\left(\frac{\ell }{2}+1\right)-\psi ^{(1)}\left(\frac{\ell +1}{2}\right)\right)+\pi ^2\right)}{\ell ^2 (\ell +1)^2}+\frac{8 \left(265 \ell ^4+280 \ell ^3-36 \ell ^2-39 \ell +33\right)}{\ell ^3 (\ell +1)^3}-65\right)\nn &+\frac{1377 \left(16 (\ell +1) \ell  H_{\ell }+\ell ^4+2 \ell ^3-39 \ell ^2-16 \ell +12\right)}{\ell ^2 (\ell +1)^2}+\frac{(2592 \zeta (3)-1865) (\ell -2) (\ell +3)}{\ell  (\ell +1)}\bigg)\nn
&+O(\e^5)\,,
\end{align}
where $H_{\ell }^{(2)}$ is the generalized harmonic number of power 2 and $\psi ^{(1)}(x)$ is the polygamma function. 
 Now we take the large spin limit of \eqref{gphi2} and compare it with \eqref{gphi}. This results in the following solution for the OPE coefficients of $\f^2$,
\be
C_0^{(0)}=2, \quad C_0^{(1)}=-\frac{2}{3}\, , \quad  C_0^{(2)}=-\frac{34}{81}\,.
\ee
This gives an alternative way of computing the OPE coefficients of $\f^2$ using the known results from bootstrap and Feynman diagrams.
\section{$\phi^3$ theory  with Mellin bootstrap}\label{ep6d}

In this section we will obtain the OPE data at large spin, for the $\phi^3$ theory in $d=6-\e$ dimension and compare them with the general spin OPE data, that can be obtained using Mellin bootstrap techniques \cite{uslong}. In this theory there is a fundamental scalar $\phi$ with conformal dimension,
\be\label{phi3phi}
\D_\phi=2 - \frac{5}{9}\e+O(\e^2)\,.
\ee
In the OPE of $\phi\times\phi$ the operator spectrum contains the operator $\phi$ itself with the OPE coefficient \cite{uslong},
\be\label{opephi}
C_{\ell_m=0}=-\frac{2\e}{3}\,.
\ee
There are also the double field higher spin operators $\phi\partial^\ell\phi$ with dimension $2\D_\phi+\ell+\g$. 

\subsection{OPE data at large spin}
First we will evaluate the anomalous dimension and OPE coefficients at large spin using \eqref{anmdim}. Note that the minimal twist operator here is the operator $\phi$. Using \eqref{phi3phi} with $\ell_m=0$ in \eqref{anmdim} one gets,
\be
\g_0=\frac{4\e}{3\ell^2}\,.
\ee
Similarly using \eqref{ope} one gets the large spin OPE coefficients,
\be\label{ope3}
\delta C_0=-\frac{4\e}{3\ell^2}\log 2\,.
\ee
We can also obtain the terms subleading in $1/\ell$\,. They are given by,
\be
\frac{\g_1}{\ell}=-\frac{4\e}{\ell^3}\,,\hspace{2cm} \frac{\d C_1}{\ell}=\frac{\e(1+6 \log 4)}{\ell^3}\,.
\ee
To compute the subleading terms with  $1/\ell^4$ suppression we have to incorporate the contribution of the higher spin operators themselves. This is easy to see since their twists start with $4+O(\e)$.

We have only computed the $O(\e)$ term. Before we discuss the $O(\e^2)$ order, we will compute the general spin anomalous dimensions and OPE coefficients using Mellin Bootstrap. Then we will discuss how to use that information to obtain $O(\e^2)$ data, and also discuss the possible difficulties. 

\subsection{A short review of Mellin bootstrap }\label{anmope}
Let us start with a quick review of the Bootstrap in Mellin space, introduced in \cite{prl, uslong, booton}. The idea is to write a 4-point of scalars as the sum over Witten diagrams. For identical scalars we have,
\be
(x_{12}^2x_{34}^2)^{2\D_\phi}\langle \phi(x_1)\phi(x_2)\phi(x_3)\phi(x_4) \rangle=\mathcal{A}(u,v)=\sum_{\Delta, \ell} c_{\Delta, \ell}\bigg( W_{\Delta,\ell}^{(s)}(u,v)+W_{\Delta,\ell}^{(t)}(u,v)+W_{\Delta,\ell}^{(u)}(u,v) \bigg)\,.
\ee
Here $W_{\Delta,\ell}^{(s,t,u)}(u,v)$ are the $s,t$ and $u$ channel Witten diagrams respectively. They are conveniently written in Mellin space as,
\be\label{mellinInt}
W_{\Delta,\ell}=\int_{-i\infty}^{i\infty} \frac{ds dt}{(2\pi i)^2}  \ u^s  v^t \G^2(s+t)\G^2(-t)\G^2(\D_\phi-s) M_{\D,\ell}(s,t)\,.
\ee
Here $M_{\D,\ell}(s, t)$ is called the Mellin amplitude of the Witten diagram $W_{\D,\ell}(u, v)$. In the $s$-channel it is given by,
\begin{align}
\begin{split}\label{sunitrymell}
M^{(s)}_{\D,\ell}(s, t) = \int_{-i\infty}^{i\infty} d\nu\, \mu^{(s)}_{\Delta,\ell}(\nu)
\Omega_{\nu, \ell}^{(s)}(s)P^{(s)}_{\nu, \ell}(s,t)
\end{split}
\end{align}
where the notations are given by,
\be\label{specunitry}
\mu^{(s)}_{\Delta,\ell}(\nu)=\frac{\G^2(\frac{2\D_\phi-h+\ell+\nu}{2})\G^2(\frac{2\D_\phi-h+\ell-\nu}{2})}{2\pi i((\D-h)^2-\nu^2)\G(\nu)\G(-\nu)(h+\nu-1)_\ell(h-\nu-1)_\ell}\,.
\ee
and 
\be
\Omega_{\nu, \ell}^{(s)}(s)=\frac{\G(\lambda_2-s)\G(\bar{\lambda}_2-s)}{\G(\D_\phi-s)^2}\,.
\ee
Here $P^{(s)}_{\Delta-h,\ell}(s,t)$ is a Mack polynomial of degree $\ell$. Their form is shown explicitly in Appendix \ref{mack}. Also $\l_2=(h+\nu -\ell)/2$ and $\bar{\l}_2  =(h- \nu -\ell)/2$. The $t$ and $u$ channel Mellin amplitudes are obtained by interchanging $(s\to t+\D_\phi,t\to s-\D_\phi)$ and $(s\to\D_\phi-s-t)$ respectively. 

The operator content of the OPE comes from the poles of the Mellin amplitude $M_{\D,\ell}(s,t)$. In order to have the correct $u$, $v$ dependencies of a certain channel, the Mellin amplitude must have certain poles. In particular for the $s$-channel OPE we require poles at,
\be
2s=(\D-\ell)+2 n \text{  where  } n=0,1,2\cdots\,,
\ee
These poles are called the physical poles.

There are also poles that come from the measure of the Mellin integral \eqref{mellinInt}. Such poles occur at,
\be\label{unphyspoles}
s=\D_\phi+n \text{  where  } n=0,1,2\cdots\,.
\ee
These poles do not correspond to operators present in the OPE, and hence they are called unphysical poles. The idea of Mellin bootstrap is to equate their residues to 0. These equations can be summarized as,
\be\label{dblres}
\sum_{\D \neq 0} (c_{\D,\ell} q^{(2,s)}_{\D, \ell} +2\sum_{\ell'}c_{\D,\ell'}q^{(2,t)}_{\D, \ell |\ell'})=0
\ee
and 
\be\label{snglres}
2q^{(1,t)}_{\D=0, \ell |0}+\sum_{\D \neq 0}(c_{\D,\ell} q^{(1,s)}_{\D, \ell}+2\sum_{\ell'}c_{\D,\ell'}q^{(1,t)}_{\D, \ell |\ell'}) =0\,.
\ee
The notations are defined in Appendix \ref{not}. Here we point out that $q^{(1,s)},q^{(2,s)}$ are $s$-channel contributions and $q^{(1,t)},q^{(2,t)}$ denote the crossed channel contributions. The two different equations together determine the anomalous dimensions and OPE coefficients of operators in the $s$-channel.


\subsection{Higher spin OPE data}
Now let us use the above equations to for the higher spin operators in $\phi^3$ theory. As an input we will use the dimension and OPE coefficient of $\phi$ given by \eqref{phi3phi} and \eqref{opephi} respectively. Let us write the unknowns as 
\be\label{phi3g}
\g=\d_{\ell}^{(1)} \e+O(\e^2)\hspace{1cm}\text{and} \hspace{1cm} C_\ell=C_\ell^{(0)}+\e C_{\ell}^{(1)}+O(\e^2)\,.
\ee
Using this in \eqref{dblres} we obtain for the $s$-channel,
\be
c_{\D,\ell}q^{(2,s)}_{\D,\ell}=\frac{2^{-1-\ell } C^{(0)}_{\ell } (3+2 \ell ) \d_{\ell}^{(1)}   \Gamma ^2(3+2 \ell )}{\Gamma ^4(2+\ell ) \Gamma (3+\ell )} \epsilon \ + \ O(\e^2) \,,
\ee
and in the crossed channels we get,
\be\label{phi3t1}
c_{\D,\ell'=0}q^{(2,t)}_{\D,\ell|\ell'=0}=\frac{(2)^{-\ell }\text{  }\Gamma (4+2 \ell )}{\Gamma ^2(2+\ell ) \Gamma (3+\ell )}\left(-\frac{2}{3}\epsilon \right)+O\left(\epsilon ^2\right)\,.
\ee
Again in the second equation \eqref{snglres} we get from the $s$-channel,
\begin{align}
&c_{\D,\ell}q^{(1,s)}_{\D,\ell} =\frac{ 2^{4+3 \ell } C_{\ell }^{(0)} (3+2 \ell ) \Gamma ^2\left(\frac{3}{2}+\ell \right)}{\pi  \Gamma ^2(2+\ell ) \Gamma (3+\ell )}+\e\, \frac{2^{4+3 \ell }   \Gamma ^2\left(\frac{3}{2}+\ell \right)}{\pi  \Gamma ^2(2+\ell ) \Gamma (3+\ell )}\left(C_{\ell }^{(0)}(3+2 \ell ) \left(-\frac{10}{9}+ \d_{\ell}^{(1)} \right) \big(\log (4)\right.\nn & \left.+H_{\frac{1}{2}+\ell }-2 H_{1+\ell }\big)+\frac{C_{\ell }^{(1)} (2+\ell ) (3+2 \ell )-C_{\ell }^{(0)} \left((1+\ell ) \d_{\ell}^{(1)}-\frac{10}{9}\left((2+\ell ) (3+2 \ell ) \gamma_E-1-\ell \right)\right)}{2+\ell }\right) + O(\e^2)\,.
\end{align}
and in the crossed channels,
\begin{align}\label{phi3t2}
c_{\D,\ell'}q^{(1,t)}_{\D,\ell|\ell'=0} & =\sum_{q=0}^{\ell} \frac{(-1)^{-q+1} 2^{1-\ell }\Gamma (3+q+\ell ) \Gamma (3+2 \ell ) \ \e }{3 \Gamma (2+q) \Gamma (3+q) \Gamma ^2(2+\ell ) \Gamma (3+\ell ) \Gamma (1-q+\ell )}\bigg(\frac{q-2}{1+q}+\frac{2 q \ell }{1+q}+\frac{2}{2+\ell }\nonumber\\&
-2 (3+2 \ell ) H_{2+q}-4 (3+2 \ell ) H_{1+\ell }+2 (3+2 \ell ) H_{2+q+\ell }+2 (3+2 \ell ) H_{2+2 \ell }\bigg)+ O(\e^2)\,,
\end{align}
and finally the disconnected piece,
\be
q^{(1,t)}_{0,\ell|0}=-\frac{2^{1-\ell } \Gamma (4+2 \ell )}{\ell ! \Gamma ^2(2+\ell )}+\frac{5 \times 2^{2-\ell }  \Gamma (4+2 \ell ) \left(\gamma _E-\psi (2+\ell )+\psi (4+2 \ell )-1\right)}{9 \ell ! \Gamma ^2(2+\ell )} \ \epsilon \ + \ O(\e^2)\,.
\ee
With the above we can solve \eqref{dblres} and \eqref{snglres} to obtain the anomalous dimension,
\be
\g=\frac{4 \ \e}{3(\ell+2) (\ell+1)} \ + O(\e^2)\,.
\ee
and the OPE coefficients,
\be
C_\ell=\frac{2 \Gamma ^2(2+\ell ) \Gamma (3+\ell )}{\ell ! \Gamma (3+2 \ell )} \ + \ C_{\ell}^{(1)} \e +O(\e^2)\,.
\ee
The $C_{\ell}^{(1)}$ can be expressed as a finite sum and obtained for any $\ell$. For the first few spins, they are given by,
\be
C_2^{\text{(1)}}=-\frac{317}{450}\,,  \ C_4^{\text{(1)}}=-\frac{8717}{39690}\,,  \ C_6^{\text{(1)}}=-\frac{54146753}{1391349960}\,,  \ C_8^{\text{(1)}}=-\frac{29214937}{5584724145}\,,  \ C_{10}^{\text{(1)}}=-\frac{1364174411}{2280823905360}\,.
\ee

Let us now elaborate on why it was simple to obtain the $O(\e)$ results and what comes in the way for the next order. First note that in the crossed channels, \eqref{phi3t1} and \eqref{phi3t2} only the spin 0 exchange i.e the $\phi$ contributes at $O(\e)$. The other operators such as the higher spin operators, or operators with twists greater than $2\D_\phi$ contribute from $O(\e^2)$ or beyond. Even in the $s$-channel the higher twist operators contribute from a higher order. Thus we easily obtain the $O(\e)$ results. 

In section 5 of \cite{uslong}  the difficulties in going beyond $O(\e^3)$ for the $\phi^4$ in $4-\e$ dimension are described. The problems one faces in getting $O(\e^2)$ for the $\phi^3$ theory in $6-\e$ dimension are similar to the ones faced in getting $O(\e^4)$ of the $\phi^4$ theory. There are two main problems at these higher orders. One is the involvement of infinitely many operators as discussed above. The other problem is the infinite number of poles of $\nu$ that can contribute. In \eqref{ts} there are only two poles that contribute at $O(\e)$ in the crossed channels, which makes the calculation simple. The former problem is one that is related to the intrinsic difficulty that one expects at higher orders in perturbation. However the latter is a calculational hurdle, that can supposedly be bypassed. In the following section, we demonstrate a calculation that does precisely that, by writing the Witten diagrams differently and avoiding the sum over infinite poles of $\nu$.

\subsection{A higher loop calculation}

We have seen in section \ref{epexp} that in lightcone bootstrap higher orders in perturbation are obtained from a sum over contributions from infinite number of operators. In lightcone bootstrap, for the $\phi^3$ theory in $d=6-\e$, the higher spin operators of the type $O_\ell \sim \phi \partial_{\mu_1}\partial_{\mu_2}\cdots \phi$ contribute from $O(\e^2)$ in the crossed channel.  So it is not surprising to expect the same from Mellin bootstrap, and here too these operators contribute from $O(\e^2)$ in the $t$ and $u$ channels. In this section we will demonstrate a calculation that systematically computes these contributions in the crossed channels. For simplicity we will take only the spin 0 operator in the $s$-channel, and the calculation would correspond to its OPE coefficient at $O(\e^2)$ order.

Instead of \eqref{sunitrymell} we will use the following expression for the Mellin amplitude of the Witten diagram \cite{uslong,regge},
\be\label{mellwit}
M_{\Delta,\ell}^{(s)}(s,t)=\sum_{m=0}^{\infty} \frac{(-1)^m\sin{\pi(\Delta-h)}\G(h-\Delta-m)}{m!\G(\frac{\Delta_1+\Delta_2-\Delta+\ell-2m}{2}) \G(\frac{\Delta_3+\Delta_4-\Delta+\ell-2m}{2})}\frac{Q^{\Delta}_{\ell,m}(t)}{2s-\Delta+\ell-2m}+R_{\ell-1}(s,t) \,.
\ee
The $Q^{\Delta}_{\ell,m}(t)$ is defined in Appendix \ref{mack}. The last term is a polynomial ambiguity, which is present in the definition of the exchange Witten diagram. It comes from how one chooses the scalar-scalar-spin vertex. In the calculations of this section we will simply drop it and come back to it in the end.

We will put this Mellin amplitude in \eqref{mellinInt} and compute the coefficient of $u^{\D_\phi}\log u (1-v)^0$. This term will get contribution from only the spin 0 operators. In all the three channels, we will simply put $v=1$ in order to have this particular term. Note that this is essentially same as expanding in terms of $Q_\ell(t)$, since we are looking at spin 0 in $s$-channel.

Taking the residue at $s=\D_\phi$ in \eqref{mellinInt} and getting the coefficient of the $\log$ term from the $s$-channel we get,
\be
\mathcal{A}^{(s)}(u,v)\big|_{u^{\D_\phi}\log u}=\int \frac{dt}{2\pi i} v^t \bigg( -\sum_{m=0}^\infty\frac{2 C_{\D,\ell}\Gamma ^2(-t) \Gamma (\Delta ) \Gamma (1-h+\Delta ) \Gamma ^2\left(t+\Delta _{\phi }\right)}{m! \left(2 m+\Delta -2 \Delta _{\phi }\right) \Gamma ^4\left(\frac{\Delta }{2}\right) \Gamma (1-h+m+\Delta ) \Gamma ^2\left(-m-\frac{\Delta }{2}+\Delta _{\phi }\right)} \bigg)  \,.
\ee
For the $t$-integral we simply put $v=1$ and use Barnes Lemma. Then carrying out the sum over $m$ we get,
\be
\mathcal{A}^{(s)}(u,v)\big|_{u^{\D_\phi}\log u}=-\frac{2C_{\D,\ell}\text{  }\Gamma (\Delta ) \Gamma (1-h+\Delta ) \Gamma ^4\left(\Delta _{\phi }\right)\text{  }\Gamma \left(-h+2 \Delta _{\phi }\right)}{\left(\Delta -2 \Delta _{\phi }\right) \Gamma ^4\left(\frac{\Delta }{2}\right) \Gamma \left(2 \Delta _{\phi }\right) \Gamma ^2\left(-\frac{\Delta }{2}+\Delta _{\phi }\right) \Gamma \left(-h+\frac{\Delta }{2}+\Delta _{\phi }\right) \Gamma \left(1-h+\frac{\Delta }{2}+\Delta _{\phi }\right)}\,.
\ee
Now we put $\D=\D_\phi=2 - \frac{5}{9}\e+\d_\phi^{(2)}\e^2+O(\e^3)$\, for the spin 0 $\phi$ exchange. Also with $d=2h=6-\e$ and $C_{\D,0}=C_{0}^{(0)}\e+C_{0}^{(1)}\e^2+O(\e^3)$\,, we get from the above,
\be\label{schindr}
\mathcal{A}^{(s)}(u,v)\big|_{u^{\D_\phi}\log u}= C_0^{(0)}\epsilon +\frac{9}{4} \epsilon ^2 \left(\frac{4 C_0^{(1)}}{9}+C_0^{(0)}\left(-\frac{50}{243}+6 \delta _{\phi }^{(2)}\right)\right)\,.
\ee
Now let us come to the crossed channels. We substitute $s\to\D_\phi+t, t=s-\D_\phi$ and $s\to \D_\phi-s-t$ in \eqref{mellwit} to get the $t$ and $u$ channel Mellin amplitudes respectively. Here the calculation becomes tricky if one follows the same route as in $s$-channel. This is because after these substitutions, the $t$-integral and sum over $m$ is not straightforward. So, what we do is to expand the expression in $\e$ first and then integrate over $t$ and sum over $m$. For example, in the $u$-channel we get for the spin $\ell' = 0$ exchange,
\be
\mathcal{A}^{(u)}(u,v)\bigg|_{u^{\D_\phi}\log u}^{\ell'=0}=\int \frac{dt}{2\pi i}\sum_m\bigg(\frac{2 C_{\Delta ,0} u^{\Delta _{\phi }} v^t \Gamma ^2(-t) \Gamma (\Delta ) \Gamma (1-h+\Delta ) \Gamma ^2\left(t+\Delta _{\phi }\right)}{(2 m+2 t+\Delta ) m! \Gamma ^4\left(\frac{\Delta }{2}\right) \Gamma (1-h+m+\Delta ) \Gamma ^2\left(\Delta _{\phi }-\frac{\Delta }{2}-m\right)}\bigg)\,.
\ee
Expanding the $m=0$ term from this in $\e$ we get,
\begin{align}\label{m0term}
&-\int \frac{dt}{2\pi i} \frac{C_0^{(0)} v^t \epsilon  \Gamma ^2(-t) \Gamma ^2(2+t)}{1+t}-\nonumber \\ &\frac{v^t \epsilon ^2 \Gamma ^2(-t) \Gamma ^2(2+t)\left(18 C_0^{(1)} (1+t)+C_0^{(0)}\left(10-5(3+2 t) -20 (1+t) \gamma _E\right)-20 C_0^{(0)}(1+t) \psi (2+t)\right)}{18 (1+t)^2}\,.
\end{align}
The $m>0$ terms give
\be\label{mgrt0}
\int \frac{dt}{2\pi i}\sum_{m=1}^{\infty}  \frac{25\text{  }C_{0}^{(0)} v^t \epsilon^2 \text{  }\Gamma ^2(-t) \Gamma ^2(2+t)}{162 (m-t-1) m }\,.
\ee
The $m=0$ term above is the only term that contributes at $O(\e)$\,. The $t$ channel gives an equal contribution. Carrying out Barnes Lemma for the $O(\e)$ part of \eqref{m0term} we get from the crossed channels,
\be
\mathcal{A}^{(crossed)}(u,v)\big|_{u^{\D_\phi}\log u}= -C_0^{(0)}\epsilon  \,.
\ee
This precisely cancels the $O(\e)$ term of the $s$-channel \eqref{schindr}. The $O(\e)$ anomalous dimension of $\phi$ is hence consistent with this alternative approach. Let us now go one step further and look at the $O(\e^2)$ terms. This order gets contribution from the $m=0$ of \eqref{m0term} as well as the $m>0$ terms as shown in \eqref{mgrt0}. The latter can be summed to give, 
\be
\mathcal{A}^{(u)}(u,v)\bigg|_{u^{\D_\phi}\log u}^{\ell'=0,m>0}=-\int \frac{dt}{2\pi i} \frac{ v^t 25  \Gamma ^2(-t) \Gamma ^2(2+t) \left(\gamma _E+\psi (-t)\right)}{243(1+t) (1+2 \text{$\delta \phi $1})} \ \epsilon^2 = -\frac{25 \epsilon ^2}{108}\,.
\ee
The last step has been done putting $v=1$ and then using Barnes Lemma to integrate over $t$. 

However there are more contributions to $O(\e^2)$, and these come from the higher spin operators in crossed channels. We will now compute these contributions systematically. Taking the spin $\ell'=2$ in $u$-channel and following the same route as above we get,
\be
\mathcal{A}^{(u)}(u,v)\bigg|_{u^{\D_\phi}\log u}^{\ell'=2}=-\frac{35 C_2^{(0)} \left(6+m^2+5 m (1+t)+5 t (2+t)\right)\text{  }v^t \epsilon ^2\text{  }\left(\delta_2^{(1)}\right)^2}{4 (2+m+t) (3+m)(2+m)(1+m)}\,.
\ee
Here $\delta_2^{(1)}$ is the $O(\e)$ anomalous dimension of the spin 2 \eqref{phi3g}. Summing over $m$ and followed by integrating $t$ we get,
\be
\mathcal{A}^{(u)}(u,v)\bigg|_{u^{\D_\phi}\log u}^{\ell'=2}=-\frac{385}{96}C_{2}^{(0)}\left(\delta_2^{(1)}\right)^2\e^2\,.
\ee
These steps can be repeated for $\ell'=4,6,\cdots$ giving the general result,
\be
\mathcal{A}^{(u)}(u,v)\bigg|_{u^{\D_\phi}\log u}^{\ell'}=-\frac{3+2 \ell' }{8 \left(2+3 \ell' +{\ell'} ^2\right)^2}
\ee
This sum over even spins $\ell'$ can be carried out and it gives $\frac{1}{32} \left(3-\frac{\pi ^2}{3}\right)$\,.

The above demonstrates an example of how one would calculate contributions from other operators in the crossed channels. For the $\phi^3$ in $6-\e$ such contributions come at $O(\e^2)$ from only the higher spin operators bilinear in $\phi$\,. Now, we leave the result undetermined because of the presence of the polynomial ambiguity of \eqref{mellwit}, which we cannot fix. So, our calculation will possibly give a part of the correct result. Let us now say a few words on what role the polynomial pieces can play.

In \cite{prl,uslong} it has been commented that this piece might be important to fix in order to make the Witten diagrams a convergent basis, which is important for numerical analysis. The polynomial ambiguity might also contribute to higher orders in perturbation, which in our case is the $O(\e^2)$. In \cite{usnew}  it was shown that they will contribute from the order of $O(\g_\ell^2)$ which is indeed $O(\e^2)$ for this case (and $O(\e^4)$ in Wilson-Fisher $d=4-\e$). So in order to get the complete results at higher orders one needs to fix these ambiguities or find an alternative way to deal with them \cite{ASnew}. It is yet unknown how to do this. We leave this problem for future work.

\section{Discussion}\label{disc}
%
%
%
We analysed the higher spin operators at the Wilson-Fisher fixed point for $\phi^4$ theory in $4-\e$ dimensions and $\f^3$ theory in $6-\e$ diemnsions.  

For the former we have employed a hybrid method- we have done a large spin analysis but at higher orders of perturbation. In doing so we used results from both the newly introduced Mellin Bootstrap and also from Feynman diagram literature. The higher order terms hence obtained are unknown in literature, and moreover the technique also provides a cross-check for the results of Mellin Bootstrap. 

For the latter we have used the Mellin bootstrap technique. We calculated OPE data for general spin operators. We have also demonstrated how one can approach a higher perturbative order in this approach, that involves contributions from infinite operators in the crossed channels. 

There are several interesting future directions one can pursue:
\begin{itemize}
	
	\item 
	One can use these techniques for other theories too, like large $N$ CFTs or theories in other dimensions.	
	
	\item
	It would be interesting to systematically extend the ideas to higher orders in $\e$ and $\ell^{-1}$. This would require knowledge of higher twist operators too.
	
	\item 
	This brings us to the question of obtaining higher twist OPE data from bootstrap. One expects to get these from other kinds of correlators. It is an interesting open problem how to use Mellin bootstrap for other kinds of correlators.

	\item
	Finally it is important to understand the role of the polynomial ambiguities in Witten diagrams \cite{ASnew}. Developing the systematics of the epsilon expansion in usual bootstrap is one of the most exciting future directions. It may shed light to the polynomial ambiguities of the Witten diagram basis and can be used to fix the ambiguity following \cite{usnew}.

\end{itemize}
\section*{Acknowledgments}
We acknowledge useful discussions with  Kausik Ghosh, Rajesh Gopakumar and Aninda Sinha. We thank Aninda Sinha for comments on the manuscript. We also thank Fernando Alday for discussions and pointing out some typos.
\appendix
\section{Details of section \ref{anmope}}\label{not}
The residues from $\G^2(\D_\phi)$ give  $u^{\D_\phi+n}\log u$ and $u^{\D_\phi+n}$  dependence which are unphysical because they do not occur in the $s$-channel OPE. The residues can be expanded in the basis of the continuous Hahn polynomials $Q^{2s+\ell}_{\ell, 0}$, in the following way,
\begin{align}\label{Mtoq}
M^{(s)}(s\to\D_\phi,t)=&  \sum_{\D, \ell}\, c_{\D,\ell}q^{(s)}_{\D, \ell} \, Q^{2\D_\phi+\ell}_{\ell, 0}(t)+\cdots\nonumber\\
M^{(t)}(s\to \D_\phi,t) 
=&  \sum_{\D, \ell,\ell'}\, c_{\D,\ell}q^{(t)}_{\D, \ell|\ell'} \, Q^{2\D_\phi+\ell}_{\ell, 0}(t)+\cdots\nonumber\\
M^{(u)}(s\to\D_\phi,t)
=&  \sum_{\D, \ell,\ell'}\, c_{\D,\ell}q^{(u)}_{\D, \ell|\ell'} \, Q^{2\D_\phi+\ell}_{\ell, 0}(t)+\cdots
\end{align}
where,
\begin{align}\label{norm}
&c_{\D,\ell}= 
C_{\D,\ell}\frac{(-2)^{\ell }  (\ell +\Delta-1 ) \Gamma (1-h+\Delta ) \Gamma ^2(\ell +\Delta-1 )}{ \Gamma (\Delta -1) \Gamma ^4\left(\frac{\ell +\Delta }{2}\right) } \Gamma^{-2} \left(\frac{\ell -\Delta +2\Delta _\f}{2}\right) 
 \Gamma^{-2}  \left(\frac{\Delta +2\Delta _\f -2 h+\ell }{2}\right)
\,.
\end{align}
The $\cdots$ denote contributions from the physical and other spurious poles. The Hahn polynomials $Q^{\Delta}_{\ell,0}(t)$ are defined in terms of the Mack polynomaials $P_{\nu,\ell}^{(s)}(s,t)$ as
\be
Q^{\Delta}_{\ell,0}(t)= \frac{4^\ell }{(\Delta-1)_{\ell}(2h-\Delta-1)_{\ell}}P_{\Delta-h,\ell}(s=\frac{\Delta-\ell}{2},t) \,.
\ee
If we Taylor expand $M(s-\D_\phi,t)$ around $s=\D_\phi$,
\be
M(s-\D_\phi,t)=M(\D_\phi,t)+(s-\D_\phi)M'(\D_\phi,t)\,.
\ee
The first term gives the logarithmic unphysical term and the second gives the nonlogarithmic one (or the power law). This can be applied to \eqref{Mtoq}. In the $s$-channel one has,
\begin{align}\label{ss1}
&  q^{(s)}_{\D, \ell}(s)= -4^{1-\ell}\,\frac{(2s+\ell-1)_{\ell}\,(2h-2s-\ell-1)_{\ell}\,\G(h-\ell-2s)\,\m^{(s)}_{\D,\ell}(\nu)}{\G(\D_\f-s)^2}\,.
\end{align}
If we write this as,
\begin{align}\label{ss}
&q^{i,(s)}_{\D, \ell}(s)=q_{\D,\ell}^{(2,t)}+(s-\D_\phi)q^{(1,s)}_{\D,\ell}+O((s-\D_\phi)^2)\nonumber\\
&=-\frac{4^{1-\ell}\G(2\D_\phi+\ell-h)}{(\ell-\D+2\D_\phi)(\ell+\D+2\D_\phi-2h)}+(s-\D_\phi)\frac{4^{2-\ell}\G(2\D_\phi+\ell-h+1)}{(\ell-\D+2\D_\phi)^2(\ell+\D+2\D_\phi-2h)^2}\,,
\end{align}
the first term in the second line, is related to the $u^{\D_\phi}\log u$ term\,, and the second term is related to the coefficient of the power law term $u^{\D_\phi}$. 

The continuous Hahn polynomials $Q_{\ell,0}^\D(t)$ are orthogonal polnomials, 
and their properties  are detailed in Appendix \ref{mack}. 
We can use this in the crossed channels to get,
\begin{align}\label{ts}
q^{i,(t)}_{\D, \ell|\ell'}(s)=& \frac{1}{\kappa_{\ell}(s)} \int \frac{dt}{2\pi i}\, \G(s+t)^2\,\, Q^{2s+\ell}_{\ell, 0}(t)  \int \, d\nu\, \, \G(\la_2-t-\D_\f)\,\G(\bar{\la}_2-t-\D_\f)\,\, \m^{(t)}_{\D,\ell}(\nu)\, P^{(t)}_{\nu,\ell'}(s-\D_\f, t+\D_\f)\\
\label{tu}
q^{i,(u)}_{\D, \ell|\ell'}(s)=& \frac{1}{\kappa_{\ell}(s)}\, \int \frac{dt}{2\pi i}\, \G(s+t)^2\, \, Q^{2s+\ell}_{\ell, 0}(t)   \times  \int \, d\nu\, \, \G(\la_2-t-\D_\f)\,\G(\bar{\la}_2-t-\D_\f)\,\, \m^{(u)}_{\ell^\prime}(\nu)\, P^{(u)}_{\nu,\ell'}(s-\D_\f, t)\,.
\end{align}
We sum up the coefficients of the log and the power law terms individually from all the three channels and for each $\ell$ equate them to 0. 
\section{Mack polynomial and continuous Hahn polynomial}\label{mack}
The Mack polynomials are given by\cite{mack1, regge, dolan},
\be\label{Phat}
{P}_{\nu, \ell}(s, t) = (h+\nu-1)_{\ell}\, (h-\nu-1)_{\ell} \,\sum_{m=0}^{\ell}\sum_{n=0}^{ \ell-m}\mu^{(\ell)}_{m, n}\,\left(\frac{h+\nu-\ell}{2}-s\right)_m\,(-t)_n\,,
\ee
where, 
\begin{eqnarray}\label{mudef}
\mu_{m,n}^{(\ell)}&=&2^{-\ell} \frac{(-1)^{m+n}\ell!}{m! n! (\ell-m-n)!}(\frac{\D+\ell}{2}-m)_m (\frac{\tau}{2}+n)_{\ell-n} (\frac{\tau}{2}+m+n)_{\ell-m-n}(\ell+h-1)_{-m}(\ell+\D-1)_{n-\ell}  \nonumber \\
&\times& {}_4F_3[-m,1-h+\frac{\tau}{2},1-h+\frac{\tau}{2},n-1+\D;2-2h+\tau,\frac{\D+\ell}{2}-m,\frac{\tau}{2}+n;1]
\end{eqnarray}
and $h+\nu=\D$. 

The continuous Hahn polynomials are defined as \cite{mack1, regge, dolan, AAR},
\be\label{Qdefn}
{Q}^{2s+\ell}_{\ell,0}(t)= \frac{2^{\ell}\,((s)_{\ell})^2}{(2s+\ell-1)_{\ell}}\,{}_3F_2\bigg[\begin{matrix} -\ell,\, 2s+\ell-1,\,s+t\\
	\ \ s \ \ , \ \ \ \ \ \  s
\end{matrix};1\bigg]\,.
\ee
These are orthogonal polynomials which satisfy the following orthogonality condition ,
\be\label{ortho}
\frac{1}{2\pi i}\int_{-i\infty}^{i\infty} dt \ \G^2(s+t)\G^2(-t) {Q}^{2s+\ell}_{\ell,0}(t) {Q}^{2s+\ell '}_{\ell',0}(t)=(-1)^\ell {\kappa}_{\ell}(s)\d_{\ell,\ell'}\,,
\ee
where,
\be\label{kappadef}
{\kappa}_{\ell}(s)= \frac{4^\ell \ell!}{(2s+\ell-1)_\ell^2}\frac{\G^4(\ell+s)}{(2s+2\ell-1)\G(2s+\ell-1)}\,.
\ee
The more general $Q^{\tau+\ell}_{\ell,m}(t)$  are given by,
\be
Q^{\tau+\ell}_{\ell,m}(t)= \frac{4^\ell }{(\tau+\ell-1)_{\ell}(2h-\ta-\ell-1)_{\ell}}P_{\tau+\ell-h,\ell}(s=\frac{\tau}{2}+m,t) \,.
\ee


\begin{thebibliography}{99}


\bibitem{migdal} 
A.~A.~Migdal,
``Conformal invariance and bootstrap,''
Phys.\ Lett.\  {\bf 37B}, 386 (1971).

\bibitem{fer} 
S.~Ferrara, A.~F.~Grillo, G.~Parisi and R.~Gatto,
``Covariant expansion of the conformal four-point function,''
Nucl.\ Phys.\ B {\bf 49}, 77 (1972)
Erratum: [Nucl.\ Phys.\ B {\bf 53}, 643 (1973)].\\
S.~Ferrara, A.~F.~Grillo and R.~Gatto,
``Tensor representations of conformal algebra and conformally covariant operator product expansion,''
Annals Phys.\  {\bf 76}, 161 (1973).

\bibitem{polya}
A.~M.~Polyakov,
``Nonhamiltonian approach to conformal quantum field theory,''
Zh.\ Eksp.\ Teor.\ Fiz.\  {\bf 66}, 23 (1974).
\bibitem{bpz} 
A.~A.~Belavin, A.~M.~Polyakov and A.~B.~Zamolodchikov,
``Infinite Conformal Symmetry in Two-Dimensional Quantum Field Theory,''
Nucl.\ Phys.\ B {\bf 241}, 333 (1984). 

\bibitem{rrtv} 
  R.~Rattazzi, V.~S.~Rychkov, E.~Tonni and A.~Vichi,
  ``Bounding scalar operator dimensions in 4D CFT,''
  JHEP {\bf 0812}, 031 (2008)
  [arXiv:0807.0004 [hep-th]].

\bibitem{3dising}
S.~El-Showk, M.~F.~Paulos, D.~Poland, S.~Rychkov, D.~Simmons-Duffin and A.~Vichi,
``Solving the 3D Ising Model with the Conformal Bootstrap,''
Phys.\ Rev.\ D {\bf 86}, 025022 (2012)
[arXiv:1203.6064 [hep-th]].\\
S.~El-Showk, M.~F.~Paulos, D.~Poland, S.~Rychkov, D.~Simmons-Duffin and A.~Vichi,
``Solving the 3d Ising Model with the Conformal Bootstrap II. c-Minimization and Precise Critical Exponents,''
J.\ Stat.\ Phys.\  {\bf 157}, 869 (2014)
[arXiv:1403.4545 [hep-th]].

\bibitem{mostprecise}
F.~Kos, D.~Poland, D.~Simmons-Duffin and A.~Vichi,
``Precision islands in the Ising and O(N) models,''
JHEP {\bf 1608}, 036 (2016)
[arXiv:1603.04436 [hep-th]].


\bibitem{others}
A.~Liam Fitzpatrick, J.~Kaplan, M.~T.~Walters and J.~Wang,
``Eikonalization of conformal blocks,"
JHEP {\bf1509} (2015) 019 
arXiv:1504.01737[hep-th].\\
S.~Hellerman, D.~Orlando, S.~Reffert and M.~Watanabe,
``On the CFT operator spectrum at large global charge,"
arXiv: 1505.01537[hep-th].\\
T.~Hartman, S.~Jain and S.~Kundu,
``Causality Constraints in Conformal Field Theory,"
arXiv: 1509.00014[hep-th].\\
D.~M.~Hofman, D.~Li, D.~Meltzer, D.~Poland and F.~Rejon-Barrera,
``A Proof of the Conformal Collider Bounds,''
JHEP {\bf 1606}, 111 (2016)
[arXiv:1603.03771 [hep-th]].\\   
D.~Li, D.~Meltzer and D.~Poland,
``Conformal Collider Physics from the Lightcone Bootstrap,''
JHEP {\bf 1602}, 143 (2016)
[arXiv:1511.08025 [hep-th]].\\
L.~Alvarez-Gaume, O.~Loukas, D.~Orlando and S.~Reffert,
``Compensating strong coupling with large charge,''
arXiv:1610.04495 [hep-th].
\bibitem{spins}
A.~Castedo Echeverri, E.~Elkhidir, D.~Karateev and M.~Serone,
``Deconstructing Conformal Blocks in 4D CFT,''
JHEP {\bf 1508}, 101 (2015)
[arXiv:1505.03750 [hep-th]].\\
L.~Iliesiu, F.~Kos, D.~Poland, S.~S.~Pufu, D.~Simmons-Duffin and R.~Yacoby,
``Bootstrapping 3D Fermions,''
JHEP {\bf 1603}, 120 (2016)
[arXiv:1508.00012 [hep-th]].\\
L.~Iliesiu, F.~Kos, D.~Poland, S.~S.~Pufu, D.~Simmons-Duffin and R.~Yacoby,
``Fermion-Scalar Conformal Blocks,''
JHEP {\bf 1604}, 074 (2016)
[arXiv:1511.01497 [hep-th]].\\
A.~Castedo Echeverri, E.~Elkhidir, D.~Karateev and M.~Serone,
``Seed Conformal Blocks in 4D CFT,''
JHEP {\bf 1602}, 183 (2016)
[arXiv:1601.05325 [hep-th]].\\
S.~Giombi, V.~Kirilin and E.~Skvortsov,
``Notes on Spinning Operators in Fermionic CFT,''
JHEP {\bf 1705}, 041 (2017)
[arXiv:1701.06997 [hep-th]]
\bibitem{susy} 
C.~Beem, L.~Rastelli and B.~C.~van Rees,
``The $\mathcal N=4$ Superconformal Bootstrap,''
Phys.\ Rev.\ Lett.\  {\bf 111}, 071601 (2013)
[arXiv:1304.1803 [hep-th]].\\
C.~Beem, M.~Lemos, P.~Liendo, L.~Rastelli and B.~C.~van Rees,
``The $ \mathcal{N}=2 $ superconformal bootstrap,''
JHEP {\bf 1603}, 183 (2016)
[arXiv:1412.7541 [hep-th]].\\
M.~Lemos and P.~Liendo,
``Bootstrapping $ \mathcal{N}=2 $ chiral correlators,''
JHEP {\bf 1601}, 025 (2016)
[arXiv:1510.03866 [hep-th]].\\
A.~Bissi and T.~Łukowski,
``Revisiting $ \mathcal{N}=4 $ superconformal blocks,''
JHEP {\bf 1602}, 115 (2016)
[arXiv:1508.02391 [hep-th]].\\
C.~Beem, M.~Lemos, L.~Rastelli and B.~C.~van Rees,
``The (2, 0) superconformal bootstrap,''
Phys.\ Rev.\ D {\bf 93}, no. 2, 025016 (2016)
[arXiv:1507.05637 [hep-th]].\\
Y.~Kimura and R.~Suzuki,
``Negative anomalous dimensions in $\mathcal{N} =$ 4 SYM,''
Nucl.\ Phys.\ B {\bf 900}, 603 (2015)
[arXiv:1503.06210 [hep-th]].\\
S.~Hellerman and S.~Maeda,
``On the Large $R$-charge Expansion in ${\mathcal N} = 2$ Superconformal Field Theories,''
arXiv:1710.07336 [hep-th].

\bibitem{hogerv} 
M.~Hogervorst and B.~C.~van Rees,
``Crossing Symmetry in Alpha Space,''
arXiv:1702.08471 [hep-th].\\
D.~Karateev, P.~Kravchuk and D.~Simmons-Duffin,
``Weight Shifting Operators and Conformal Blocks,''
arXiv:1706.07813 [hep-th].\\
G.~F.~Cuomo, D.~Karateev and P.~Kravchuk,
``General Bootstrap Equations in 4D CFTs,''
arXiv:1705.05401 [hep-th].\\
A.~Codello, M.~Safari, G.~P.~Vacca and O.~Zanusso,
``Leading CFT constraints on multi-critical models in $d > 2$,''
JHEP {\bf 1704}, 127 (2017)
[arXiv:1703.04830 [hep-th]].\\
A.~Codello, M.~Safari, G.~P.~Vacca and O.~Zanusso,
``Functional perturbative RG and CFT data in the $\epsilon$-expansion,''
arXiv:1705.05558 [hep-th].\\
M.~Hogervorst,
``Crossing Kernels for Boundary and Crosscap CFTs,''
arXiv:1703.08159 [hep-th].\\
L.~Rastelli and X.~Zhou,
``The Mellin Formalism for Boundary CFT$_d$,''
arXiv:1705.05362 [hep-th].\\
S.~Giombi, V.~Gurucharan, V.~Kirilin, S.~Prakash and E.~Skvortsov,
``On the Higher-Spin Spectrum in Large N Chern-Simons Vector Models,''
JHEP {\bf 1701}, 058 (2017)
[arXiv:1610.08472 [hep-th]].\\
A.~N.~Manashov and E.~D.~Skvortsov,
``Higher-spin currents in the Gross-Neveu model at 1/n$^{2}$,''
JHEP {\bf 1701}, 132 (2017)
[arXiv:1610.06938 [hep-th]].\\
A.~Lewkowycz, G.~J.~Turiaci and H.~Verlinde,
``A CFT Perspective on Gravitational Dressing and Bulk Locality,''
JHEP {\bf 1701}, 004 (2017)
[arXiv:1608.08977 [hep-th]].\\
W.~Li,
``Inverse Bootstrapping Conformal Field Theories,''
arXiv:1706.04054 [hep-th].\\
Z.~Li and N.~Su,
``3D CFT Archipelago from Single Correlator Bootstrap,''
arXiv:1706.06960 [hep-th].\\
A.~Sever and A.~Zhiboedov,
``On Fine Structure of Strings: The Universal Correction to the Veneziano Amplitude,''
arXiv:1707.05270 [hep-th].\\
S.~Hikami,
``Conformal Bootstrap Analysis for Single and Branched Polymers,''
arXiv:1708.03072 [hep-th].\\
L.~Di Pietro and E.~Stamou,
``Scaling dimensions in QED$_3$ from the $\epsilon$-expansion,''
arXiv:1708.03740 [hep-th].\\
C.~Melby-Thompson and C.~Schmidt-Colinet,
``Double Trace Interfaces,''
arXiv:1707.03418 [hep-th].\\
H.~Isono,
``On conformal correlators and blocks with spinors in general dimensions,''
arXiv:1706.02835 [hep-th].\\
L.~Zambelli and O.~Zanusso,
``Lee-Yang model from the functional renormalization group,''
Phys.\ Rev.\ D {\bf 95}, no. 8, 085001 (2017)
[arXiv:1612.08739 [hep-th]].
\bibitem{Li2} 
F.~Kos, D.~Poland, D.~Simmons-Duffin and A.~Vichi,
``Bootstrapping the O(N) Archipelago,''
JHEP {\bf 1511}, 106 (2015)
[arXiv:1504.07997 [hep-th]].\\
D.~Li, D.~Meltzer and D.~Poland,
``Conformal Bootstrap in the Regge Limit,''
arXiv:1705.03453 [hep-th].\\
M.~Hogervorst,
``Dimensional Reduction for Conformal Blocks,''
JHEP {\bf 1609}, 017 (2016)
[arXiv:1604.08913 [hep-th]].\\
A.~L.~Fitzpatrick, J.~Kaplan, M.~T.~Walters and J.~Wang,
``Hawking from Catalan,''
JHEP {\bf 1605}, 069 (2016)
[arXiv:1510.00014 [hep-th]].\\
F.~Rejon-Barrera and D.~Robbins,
``Scalar-Vector Bootstrap,''
JHEP {\bf 1601}, 139 (2016)
[arXiv:1508.02676 [hep-th]].\\
J.~D.~Qualls,
``Universal Bounds on Operator Dimensions in General 2D Conformal Field Theories,''
arXiv:1508.00548 [hep-th].
\bibitem{gliozzi2} 
F.~Gliozzi, A.~Guerrieri, A.~C.~Petkou and C.~Wen,
``Generalized Wilson-Fisher Critical Points from the Conformal Operator Product Expansion,''
Phys.\ Rev.\ Lett.\  {\bf 118}, no. 6, 061601 (2017)
[arXiv:1611.10344 [hep-th]].\\
K.~Roumpedakis,
``Leading Order Anomalous Dimensions at the Wilson-Fisher Fixed Point from CFT,''
JHEP {\bf 1707}, 109 (2017)
[arXiv:1612.08115 [hep-th]]\\
P.~Liendo,
``Revisiting the dilatation operator of the Wilson–Fisher fixed point,''
Nucl.\ Phys.\ B {\bf 920}, 368 (2017)
[arXiv:1701.04830 [hep-th]]\\
F.~Gliozzi, A.~L.~Guerrieri, A.~C.~Petkou and C.~Wen,
``The analytic structure of conformal blocks and the generalized Wilson-Fisher fixed points,''
JHEP {\bf 1704}, 056 (2017)
[arXiv:1702.03938 [hep-th]].\\
A.~Söderberg,
``Anomalous Dimensions in the WF O($N$) Model with a Monodromy Line Defect,''
arXiv:1706.02414 [hep-th].\\
A.~N.~Manashov, E.~D.~Skvortsov and M.~Strohmaier,
``Higher spin currents in the critical $O(N$) vector model at $1/N^{2}$,''
JHEP {\bf 1708}, 106 (2017)
arXiv:1706.09256 [hep-th].\\
C.~Behan,
``Conformal manifolds: ODEs from OPEs,''
arXiv:1709.03967 [hep-th].
\bibitem{rastnew} 
L.~Rastelli and X.~Zhou,
``How to Succeed at Holographic Correlators Without Really Trying,''
arXiv:1710.05923 [hep-th].
\bibitem{zhiboedov2} 
M.~Kulaxizi, A.~Parnachev and A.~Zhiboedov,
``Bulk Phase Shift, CFT Regge Limit and Einstein Gravity,''
arXiv:1705.02934 [hep-th].

\bibitem{polandnew} 
A.~Dymarsky, F.~Kos, P.~Kravchuk, D.~Poland and D.~Simmons-Duffin,
``The 3d Stress-Tensor Bootstrap,''
arXiv:1708.05718 [hep-th].


\bibitem{tauberian} 
J.~Qiao and S.~Rychkov,
``A tauberian theorem for the conformal bootstrap,''
arXiv:1709.00008 [hep-th].
 \bibitem{heem}
 I.~Heemskerk, J.~Penedones, J.~Polchinski and J.~Sully,
 ``Holography from Conformal Field Theory,''
 JHEP {\bf 0910}, 079 (2009)
 arXiv:0907.0151 [hep-th].
 
 \bibitem{caronhout} 
 S.~Caron-Huot,
 ``Analyticity in Spin in Conformal Theories,''
 arXiv:1703.00278 [hep-th].
 
 \bibitem{sleight} 
 M.~S.~Costa, V.~Goncalves and J.~Penedones,
 ``Spinning AdS Propagators,''
 JHEP {\bf 1409}, 064 (2014)
 [arXiv:1404.5625 [hep-th]].\\
 C.~Sleight and M.~Taronna,
 ``Spinning Witten Diagrams,''
 JHEP {\bf 1706}, 100 (2017)
 [arXiv:1702.08619 [hep-th]].\\
 S.~Giombi, C.~Sleight and M.~Taronna,
 ``Spinning AdS Loop Diagrams: Two Point Functions,''
 arXiv:1708.08404 [hep-th].\\
 E.~Hijano, P.~Kraus, E.~Perlmutter and R.~Snively,
 ``Witten Diagrams Revisited: The AdS Geometry of Conformal Blocks,''
 JHEP {\bf 1601}, 146 (2016)
 [arXiv:1508.00501 [hep-th]].\\
 S.~S.~Gubser and S.~Parikh,
 ``Geodesic bulk diagrams on the Bruhat-Tits tree,''
 arXiv:1704.01149 [hep-th].\\
 A.~Castro, E.~Llabrés and F.~Rejon-Barrera,
 ``Geodesic Diagrams, Gravitational Interactions  OPE Structures,''
 JHEP {\bf 1706}, 099 (2017)
 [arXiv:1702.06128 [hep-th]].\\
 C.~Sleight and M.~Taronna,
 ``Higher spin gauge theories and bulk locality: a no-go result,''
 arXiv:1704.07859 [hep-th].

\bibitem{anboot} 
A.~L.~Fitzpatrick, J.~Kaplan, D.~Poland and D.~Simmons-Duffin,
``The Analytic Bootstrap and AdS Superhorizon Locality,''
JHEP {\bf 1312}, 004 (2013)
[arXiv:1212.3616 [hep-th]].

\bibitem{komargodski} 
Z.~Komargodski and A.~Zhiboedov,
``Convexity and Liberation at Large Spin,''
JHEP {\bf 1311}, 140 (2013)
[arXiv:1212.4103 [hep-th]].

\bibitem{alday1} 
L.~F.~Alday and A.~Zhiboedov,
``An Algebraic Approach to the Analytic Bootstrap,''
JHEP {\bf 1704}, 157 (2017)
[arXiv:1510.08091 [hep-th]].


\bibitem{alday2} 
L.~F.~Alday,
``Large Spin Perturbation Theory,''
arXiv:1611.01500 [hep-th].

\bibitem{alday3} 
L.~F.~Alday,
``Solving CFTs with Weakly Broken Higher Spin Symmetry,''
arXiv:1612.00696 [hep-th].

\bibitem{alday4} 
L.~F.~Alday, A.~Bissi and T.~Lukowski,
``Large spin systematics in CFT,''
JHEP {\bf 1511}, 101 (2015)
[arXiv:1502.07707 [hep-th]].
\bibitem{alday5} 
L.~F.~Alday and A.~Zhiboedov,
``Conformal Bootstrap With Slightly Broken Higher Spin Symmetry,''
JHEP {\bf 1606}, 091 (2016)
[arXiv:1506.04659 [hep-th]].
\bibitem{alday6} 
L.~F.~Alday and A.~Bissi,
``Crossing symmetry and Higher spin towers,''
arXiv:1603.05150 [hep-th].

\bibitem{kss}


A.~Kaviraj, K.~Sen and A.~Sinha,
``Analytic bootstrap at large spin,''
JHEP {\bf 1511}, 083 (2015)
[arXiv:1502.01437 [hep-th]].\\
A.~Kaviraj, K.~Sen and A.~Sinha,
``Universal anomalous dimensions at large spin and large twist,"
JHEP \textbf{1507} (2015) 026,
arXiv:1504.00772 [hep-th].



\bibitem{kss2}
P.~Dey, A.~Kaviraj and K.~Sen,
``More on analytic bootstrap for O(N) models,''
JHEP {\bf 1606}, 136 (2016)
[arXiv:1602.04928 [hep-th]].\\
G.~Vos,
``Generalized Additivity in Unitary Conformal Field Theories,''
Nucl.\ Phys.\ B {\bf 899}, 91 (2015)
[arXiv:1411.7941 [hep-th]].\\
D.~Li, D.~Meltzer and D.~Poland,
``Non-Abelian Binding Energies from the Lightcone Bootstrap,''
arXiv:1510.07044 [hep-th].

\bibitem{rychkovtan}
 S.~Rychkov and Z.~M.~Tan,
``The $\epsilon$-expansion from conformal field theory,''
J.\ Phys.\ A {\bf 48} (2015) no.29,  29FT01
[arXiv:1505.00963 [hep-th]].\\
P.~Basu and C.~Krishnan,
``$\epsilon$-expansions near three dimensions from conformal field theory,''
JHEP {\bf 1511}, 040 (2015)
[arXiv:1506.06616 [hep-th]].\\
S.~Ghosh, R.~K.~Gupta, K.~Jaswin and A.~A.~Nizami,
``$\epsilon$-Expansion in the Gross-Neveu model from conformal field theory,''
JHEP {\bf 1603}, 174 (2016)
[arXiv:1510.04887 [hep-th]].\\
  A.~Raju,
  ``$\epsilon$-Expansion in the Gross-Neveu CFT,''
  JHEP {\bf 1610}, 097 (2016)
  [arXiv:1510.05287 [hep-th]].
\bibitem{sen} 
K.~Sen and A.~Sinha,
``On critical exponents without Feynman diagrams,''
J.\ Phys.\ A {\bf 49}, no. 44, 445401 (2016)
[arXiv:1510.07770 [hep-th]].
\bibitem{prl} 
R.~Gopakumar, A.~Kaviraj, K.~Sen and A.~Sinha,
``Conformal Bootstrap in Mellin Space,''
Phys.\ Rev.\ Lett.\  {\bf 118}, no. 8, 081601 (2017)
[arXiv:1609.00572 [hep-th]].

\bibitem{uslong} 
R.~Gopakumar, A.~Kaviraj, K.~Sen and A.~Sinha,
``A Mellin space approach to the conformal bootstrap,''
JHEP {\bf 1705}, 027 (2017)
[arXiv:1611.08407 [hep-th]].
\bibitem{booton} 
P.~Dey, A.~Kaviraj and A.~Sinha,
``Mellin space bootstrap for global symmetry,''
JHEP {\bf 1707}, 019 (2017)
[arXiv:1612.05032 [hep-th]].
\bibitem{usnew} 
P.~Dey, K.~Ghosh and A.~Sinha,
``Simplifying large spin bootstrap in Mellin space,''
arXiv:1709.06110 [hep-th].
\bibitem{klein} 
H.~Kleinert, J.~Neu, V.~Schulte-Frohlinde, K.~G.~Chetyrkin and S.~A.~Larin,
``Five loop renormalization group functions of O(n) symmetric phi**4 theory and epsilon expansions of critical exponents up to epsilon**5,''
Phys.\ Lett.\ B {\bf 272}, 39 (1991)
Erratum: [Phys.\ Lett.\ B {\bf 319}, 545 (1993)]
doi:10.1016/0370-2693(91)91009-K, 10.1016/0370-2693(93)91768-I
[hep-th/9503230].\\
H.~Kleinert and V.~Schulte-Frohlinde,
``Exact five loop renormalization group functions of phi**4 theory with O(N) symmetric and cubic interactions: Critical exponents up to epsilon**5,''
Phys.\ Lett.\ B {\bf 342}, 284 (1995)
[cond-mat/9503038].\\
H.~Kleinert and V.~Schulte-Frohlinde,
``Critical exponents from five-loop strong coupling phi**4 theory in 4 - epsilon dimensions,''
J.\ Phys.\ A {\bf 34}, 1037 (2001)
[cond-mat/9907214].
\bibitem{phi4}
S.~E.~Derkachov, J.~A.~Gracey and A.~N.~Manashov,
``Four loop anomalous dimensions of gradient operators in phi**4 theory,''
Eur.\ Phys.\ J.\ C {\bf 2}, 569 (1998).
\bibitem{manashov2} 
  A.~N.~Manashov, E.~D.~Skvortsov and M.~Strohmaier,
  ``Higher spin currents in the critical $O(N$) vector model at $1/N^{2}$,''
  JHEP {\bf 1708}, 106 (2017)
  [arXiv:1706.09256 [hep-th]].
\bibitem{phi4klein}
H.~Kleinert and V.~Schulte-Frohlinde, ``Critical Properties of Phi4 Theories,'' World Scientific, Singapore, 2004.
 













\bibitem{regge} 
M.~S.~Costa, V.~Goncalves and J.~Penedones,
``Conformal Regge theory,''
JHEP {\bf 1212}, 091 (2012)
[arXiv:1209.4355 [hep-th]].

 \bibitem{ASnew}
     R.~Gopakumar and A.~Sinha, ``Simplifying Mellin bootstrap'', to appear.
\bibitem{mack1} 
G.~Mack,
``D-independent representation of Conformal Field Theories in D dimensions via transformation to auxiliary Dual Resonance Models. Scalar amplitudes,''
arXiv:0907.2407 [hep-th].


\bibitem{dolan} 
F.~A.~Dolan and H.~Osborn,
``Conformal Partial Waves: Further Mathematical Results,''
arXiv:1108.6194 [hep-th].
\bibitem{AAR} G.~E.~Andrews, R.~Askey and R~Roy, ``Special functions,'' Cambridge University Press, 1999.



    
    
%
    
    
    
    
    
    
    
    
   
    
    
    
    
    
    
    
    
    
    
\end{thebibliography}
\end{document}